\documentstyle[prl,aps,floats]{revtex}
\begin{document}
\twocolumn
%@twocolumnfalse\endcsname
\wideabs{
\title{A Canonical Approach to the Quantization of the Damped Harmonic 
Oscillator}
% \author{Rabin Banerjee\footnote{rabin@bose.res.in}},
 \author{Rabin Banerjee},
 
\address{S. N. Bose National Centre for Basic Sciences \\
JD Block, Sector III, Salt Lake City, Calcutta -700 098, India.
 }
\author{ Pradip Mukherjee}
 
\address{Presidency College\\
College Street, Calcutta -700 073, India.
} 
\maketitle
\begin{abstract}
{We provide a new canonical approach for studying the quantum mechanical
damped harmonic oscillator based on the doubling of degrees of freedom
approach. Explicit expressions for Lagrangians of the elementary modes 
of the problem,
characterising both forward and backward time propagations are given.
 A Hamiltonian
analysis, showing the equivalence with the Lagrangian approach, is also done.
Based on this Hamiltonian analysis, the quantization of the model is discussed.
 }\end{abstract}
}
%\maketitle
%\twocolumn
The damped harmonic oscillator (dho) problem is characterised by the breaking
of time - reversal symmetry. A direct Lagrangian formulation is problematic because
it leads to explicitly time dependent Lagrangians \cite{Ba,D}.
 The standard approach \cite{FT,C,BGPV} is to complement
the dho by its time reversed image and work with an 
effective doubled system.  
The dynamical group of symmetry of this doubled system is found to be
SU(1,1) but no unitary irreducible representation of the symmetry exists.
Time evolution leads out of the Hilbert space of states and a
satisfactory quantization can only be achieved in the framework of 
quantum field theory. This quantization procedure is based on the composite
Lagrangian of the effective system resulting from the doubling of degrees
of freedom. The lack of individual Lagrangian prescriptions
leads to problems in quantization.

We present in this paper a new method of canonical quantization of the
dho based on the doubling of degrees of freedom.
 Explicit expressions of the Lagrangians are given that
characterise dual aspects of the forward  and  backward time propagations.
   We have shown that the two cases of overdamped and (oscillatory) underdamped motions correspond to distinct regimes 
characterised by real and complex parameters, respectively, of the constituent Lagrangians. The Hamiltonians corresponding
to the complex Lagrangians are found to be pseudo - hermitian \cite{AM}. 
We have discussed the diagonalization of the complex Hamiltonians
pertaining to this regime as a generalization of the Dirac - Heisenberg
method of treating the linear harmonic oscillator. 
The breakdown of time - reversal symmetry is manifested in our analysis
by the appearence of pseudo - hermitian Hamiltonians leading to the 
time evolution of the individual modes by nonunitary operators. However,
exploiting the pseudo - hermiticity of the individual pieces, we have shown
 that well behaved states of the composite system are formed.

 We begin with a review of the problem of the damped harmonic oscillator (dho).
The equation of motion of the one - dimensional damped harmonic oscillator is
\begin{equation}
m\ddot{x} + \gamma\dot{x} + kx = 0\label{d1}
\end{equation}
The parameters $m$, $\gamma$ and $k$ are independent of time. If
the ratio
\begin{equation}
R =\frac{k}{\frac{\gamma^2}{4m}}\label{r}
\end{equation}
is greater than one,
the motion is oscillatory with exponentially decaying amplitude. Otherwise,
the motion is nonoscillatory i.e. overdamped.
Since the system (\ref{d1}) is dissipative a straightforward Lagrangian description leading to a consistent canonical quantization
is not available.
 To develop a canonical formalism we require to consider (\ref{d1}) along with its time reversed image \cite {FT}
\begin{equation}
m\ddot{y} - \gamma\dot{y} + ky = 0\label{d22}
\end{equation}
so that the composite system is conservative. The system (\ref{d1}) and
(\ref{d22}) can be derived from the Lagrangian
\begin{equation}
L = m\dot{x}\dot{y} + \frac{\gamma}{2}\left(x\dot{y} - \dot{x}y\right)
       - kxy\label{n}
\end{equation}
where $x$ is the dho coordinate and $y$ corresponds to the time - reversed
counterpart. 
Introducing the hyperbolic coordinates $x_1$ and $x_2$ \cite{BGPV} where,
\begin{equation}
x = {1\over \sqrt{2}}(x_1 + x_2); 
y = {1\over \sqrt{2}}(x_1 - x_2) \label{c3}
\end{equation}
the above Lagrangian  can be written in a compact notation as
\begin{equation}
L = {m\over 2}g_{ij}\dot{x}_i\dot{ x}_j - {\gamma\over 2}\epsilon_{ij}x_i\dot{x_j} - {k\over 2}g_{ij}x_i x_j\label{L2}
\end{equation}
where the pseudo - Eucledian metric $g_{ij}$ is given by $g_{11}$ = -$g_{22}$ = 1 and $g_{12}$ = 0.

The Lagrangian (\ref{L2}) is invariant under the SU(1,1) transformation
\begin{equation}
x_i \to x_i + \theta\sigma_{ij}x_j\label{8}
\end{equation}
where
$\sigma$ is the first Pauli matrix
and $\theta$ is an infinitesmal parameter.

The composite Lagrangian
(\ref{L2}) is analogous to the general bidimensional oscillator Lagrangian
\begin {equation}
L = {m\over 2}\dot{x_i^2} + {B\over 2}\epsilon_{ij}x_i\dot{x_j} - {1\over 2} k x_i^2\label{L3}
\end{equation}
studied recently \cite{RL} in connection with the Landau problem.
Here one exploits dual aspects of the rotation symmetry of the problem
in analysing it in terms of opposite chiralities \cite{BG}. Symmetry of (\ref{L2}) under (\ref{8}) thus offers a possibility
of analysing the composite theory in 
terms of systems having opposite chiralities w.r.t. the continuous
transformations (\ref{8}).

  Accordingly we introduce the Lagrangian doublet
\begin{equation}
L_{\pm} = \pm{\Gamma\over 2}\epsilon_{ij}x_i\dot{x}_j - {{k_{\pm}}\over 2}g_{ij}x_ix_j\label{12}
\end{equation}
which are
separately invariant under (\ref{8}). 
The Noether charges corresponding to the transformations (\ref{8}) are
\begin{equation}
C_{\pm} = \pm{{\Gamma}\over 2}g_{ij}x_i x_j
\label{16}
\end{equation}
Thus the systems (\ref{12}) have opposite "chiralities"
w.r.t. the transformation (\ref{8}) which motivates their introduction
as possible elementary forms of (\ref{L2}).

  The synthesis of $L_+$ and $L_-$ is now done by the soldering
formalism 
which has found applications in various contexts. Duality symmetric
electromagnetic actions were constructed \cite{BW}; implications in
higher dimensional bosonization were discussed \cite{BK2}; the doublet
structure in topologically massive gauge theories was revealed \cite{BK1};
a host of phenomena in two dimensions were analysed \cite{A}.
However, the analysis that is closest in spirit to the one that will be 
presented here, demonstrated the fusion of two one dimensional chiral
oscillators rotating in opposite directions, into a normal two dimensional
oscillator \cite{BG}. Indeed replacing $g_{ij}$ by $\delta_{ij}$ in equation
(\ref{12}) converts it into a doublet of chiral oscillators.

We start from a simple sum
\begin{equation}
L(y,z) = L_+(y) + L_-(z)\label{s1}
\end{equation}
and consider the gauge transformation
\begin{equation}
\delta y_i= \delta z_i = \Lambda_i(t)\label{LL5}
\end{equation}
where $\Lambda_i$ are some arbitrary functions of time. 
Under these transformations the change in $L$ is given by
\begin{eqnarray}
\delta L(y,z)& =&\delta L_+(y) + \delta L_-(z)\nonumber\\
             & =&\Lambda_i \left(J_{i}^{+}(y) + J_{i}^{-}(z)\right)\label{s2}
\end{eqnarray}
where the currents are,
\begin{equation}
J^{\pm}_i(x) 
             = \pm \Gamma\sigma_{ij}\dot{x}_j - k_{\pm}x_i\label{CN1}
\end{equation}
The idea is to iteratively modify $L(y,z)$ by suitably introducing auxiliary
variables such that the new lagrangian is invariant under the transformations
(\ref{LL5}). To this end
an auxiliary field $B_i$ transforming as (\ref{LL5}),
\begin{equation}
\delta B_i = \Lambda_i\label{LL55}
\end{equation}
is introduced and a modified lagrangian is constructed as
\begin{eqnarray}
L(y,z,B) &=& L(y,z) - B_i(J^{(+)}_i(y)+ 
              J^{(-)}_i(z))\nonumber\\
         &-&{1\over 2}(k_+ + k_-) B_iB_i\label{LL8}
\end{eqnarray}
This lagrangian is 
 now invariant under (\ref{LL5}) and (\ref{LL55}).
Since the variable $B_i$ has no independent dynamics, it is eliminated by using its equation of motion. The residual Lagrangian no longer depends on
$y$ or $z$ individually but only on the difference $y - z$. Writing this 
difference as $x$, the residual Lagrangian reproduces (\ref{L2}) with the identification
\begin{equation}
m = -{{\Gamma^2}\over{(k_+ + k_-)}},\hspace{.2cm}
\gamma={{\Gamma(k_+ - k_-)}\over{k_+ + k_-}},\hspace{.2cm}
k = {{k_+k_-}\over{(k_+ + k_-)}}\label{id1}
\end{equation}

 The essence of the soldering procedure can be understood also in the
following alternative way. Use $x_i = y_i - z_i$ in $L(y,z)$ to eliminate
$z_i$ so that
\begin{eqnarray}
L(y,x) = &-&\frac{k_+}{2}g_{ij}y_i y_j -\frac{\Gamma}{2}\epsilon_{ij}
         \left[-2y_i\dot{x}_j  + x_i\dot{x}_j\right]\nonumber\\
        &-& \frac{k_-}{2}g_{ij}\left[y_i y_j - y_i x_j - x_i y_j + x_i x_j\right]
        \label{sn5}
\end{eqnarray}
Since there is no kinetic term for $y_i$ it is really an auxiliary variable.
Eliminating $y_i$ from $L(y,x)$ by using its equation of motion
we directly arrive at (\ref{L2}) with the 
correspondence (\ref{id1}). Note that the opposite chirality of the elementary
Lagrangians are crucial in the cancellation of the time derivative of $y$
in (\ref{sn5}) which in turn is instrumental in the success of the soldering
method.

The identification (\ref{id1}) has an immediate consequence. The ratio 
(\ref{r}) is found to be,
\begin{equation}
R =\frac{k}{\frac{\gamma^2}{4m}}
= 1 - {{(k_+ + k_-)^2}\over{(k_+ - k_-)^2}}\label{28}
\end{equation}
For real $k_+$, $k_-$, the parameters identified by (\ref{id1})
correspond to an overdamped motion of the dho{\footnote{see the discussion
below (2)}}.
 Also note that to get the coefficients $m$ and $k$ to be positive we
require $k_+$ and $k_-$ to be of opposite sign, with a suitable choice of their absolute 
values. Finally, for positive $\gamma$, $\Gamma > 0$ is required. 

Now the physically more important situation is the underdamped motion of the dho where the motion is oscillatory
with decaying amplitude. Here the parameters of (\ref{L2}) are such that the ratio $R > 1$. As already observed this 
condition cannot be simulated by the identification (\ref{id1}) for real values of $k_{\pm}$. However, if $k_+$
and $k_-$ are continued to complex values so that
\begin {equation}
k_+ = \kappa\hspace{.5cm}k_- = \kappa^*\label{31}
\end{equation}
\begin{equation}
R = 1+\left({{Re\hspace{.1cm}\kappa}\over{Im\hspace{.1cm}\kappa}}\right)^2\label{32}
\end{equation}
then $R>1$,which is the required condition for oscillatory motion. Now equation (\ref{id1}) gives
\begin{equation}
m = -{{\Gamma^2}\over{2Re\hspace{.1cm}\kappa}},\hspace{.5cm}\gamma = {{i\Gamma Im\hspace{.1cm}\kappa}
\over{Re\hspace{.1cm}\kappa}},\hspace{.5cm}k=
{{|\kappa|^2}\over{2Re\hspace{.1cm}\kappa}}\label{33}
\end{equation}
Taking $\kappa$ of the form 
\begin{equation}
\kappa = \kappa_1 +i\kappa_2\label{kn}
\end {equation}
 with $\kappa_{1,2}$ positive we find that $\Gamma$ must be purely
imaginary
\begin{equation}
\Gamma = -ig,\hspace{.4cm}g> 0\label{34}
\end{equation}
so that the parameters in (\ref{33}) are positive.
Substituting (\ref{31}) and (\ref{34}) in  (\ref{12}) we get
 the elementary modes 
\begin{eqnarray}
L_+ = -i\frac{g}{2}\epsilon_{ij}x_i \dot{x_j} - \frac{\kappa}{2}g_{ij}x_ix_j\label{ln1}\\
L_- = i\frac{g}{2}\epsilon_{ij}x_i \dot{x_j} - \frac{\kappa^*}{2}g_{ij}x_ix_j\label{ln2}
\end{eqnarray}
the soldered form of which is the Lagrangian
 (\ref{L2}) pertaining to the oscillatory limit. 
Remarkably, the Lagrangians $L_{\pm}$
are now complex conjugates of each other.

 The Lagrangians (\ref{ln1}) and (\ref{ln2}) both contain informations about
forward and backward motions in time. To see this we write $L_+$ in the form
\begin{equation}
L_+ = -igx_1\dot{x_2} - \frac{\kappa}{2}(x_1^2 - x_2^2)\label{ln3}
\end{equation}
from which the Euler - Lagrange ( E - L ) equations follow as
\begin{eqnarray}
ig\dot{x_2} = -\kappa x_1\label{sn1}\\
ig\dot{x_1} = -\kappa x_2\label{sn2}
\end{eqnarray}
According to equations (\ref{33}), (\ref{kn}) and (\ref{34})
we have 
\begin{equation}
\frac{\kappa_1}{g} = \Omega\hspace{.5cm}\rm{and}
\frac{\kappa_2}{g} = \frac{\gamma}{2m}\label{v}
\end{equation}
where
\begin{equation}
\Omega = \left(\frac{1}{m}\left(k - \frac{\gamma^2}{4m}\right)\right)^{\frac{1}{2}}\label{o}
\end{equation}
The solutions to (\ref{sn1}) and (\ref{sn2}) are easy to find. Using
(\ref{v}) these solutions can be written in terms of the physical
parameters of the d.h.o. Now substituting in (\ref{c3}), we get,
\begin{eqnarray}
x = A\rm{exp}\hspace{.1cm}(-\frac{\gamma}{2m}t)exp\hspace{.1cm}
                                                  (i\Omega t)\label{sn11}\\
y = A\rm{exp}\hspace{.1cm}(\frac{\gamma}{2m}t)exp\hspace{.1cm}
                                                 (-i\Omega t)\label{sn22}
\end{eqnarray}
 Clearly, $x$ and $y$ correspond to forward and backward time propagation
with reference to the doubling of coordinates (see (\ref{d1}) and (\ref{d22})).
The same solutions also follow from $L_-$. In this connection it may be
observed that Lagrangians structurally similar with
 (\ref{ln1}) and (\ref{ln2}) was
discussed in
\cite{BGPV} as the $m \to 0$ limit of (\ref{L2}). However, it is to be
stressed that they are not quite identical because the coefficients
of (\ref{ln1}) and (\ref{ln2}) are completely different from that of the
limiting form of (\ref{L2}). This is clearly revealed by the
calculation of the friction coefficient presented in \cite{BGPV}
which comes out to be different from that of the actual damped
oscillator. 

It will be instructive to look at the problem from the Hamiltonian approach. The Hamiltonian following from (\ref{L2}) is
\begin{eqnarray}
H& =& {1\over{2m}}\left(p_1 - {{\gamma}\over 2}x_2\right)^2\nonumber\\ 
 & +& {k\over 2}x_1^2 - \frac{1}{2m}\left(p_2 + \frac{\gamma}{2}x_1\right)^2
                 - \frac{k}{2}x_2^2\label{I2}
\end{eqnarray}
where $p_1$ = $m\dot{x}_1 + \frac{\gamma}{2}x_2$, $p_2$ = $-m\dot{x}_2 - \frac{\gamma}{2}x_1$ are the canonical momentum conjugate
to $x_1$ and $x_2$. Introduce a canonical transformation from
 $(x_1,x_2;p_1,p_2)$ to $(x_+,x_-;p_+,p_-)$ where
\begin{eqnarray}
p_{\pm} =\left(\frac{\omega_{\pm}}{2m\Omega}\right)^{\frac{1}{2}}p_1 \pm i 
             \left(\frac{m\Omega\omega_{\pm}}{2}\right)^{\frac{1}{2}}x_2\nonumber\\
x_{\pm} =\left(\frac{m\Omega}{2\omega_{\pm}}\right)^{\frac{1}{2}}x_1 \pm i
             \left(\frac{1}{2m\Omega\omega_{\pm}}\right)^{\frac{1}{2}}p_2\label{ct}
\end{eqnarray}
Such transformations, though involving only real parameters, were
used in \cite{DJT}, \cite{BK1}.
Now the composite Hamiltonian diagonalises as
\begin{equation}
H = H_+ + H_-\label{hd}
\end{equation}
where,
\begin{equation}
H_{\pm} = \frac{p_{\pm}^2}{2} + \frac{\omega_{\pm}^2x_{\pm}^2}{2}\label {dh}
\end{equation}
with the frequencies $\omega_{\pm}$,  
\begin{equation}
\omega_{\pm} = \Omega \pm \frac{i\gamma}{2m}\label{freq}
\end{equation}

The Hamiltonians $H_{\pm}$ can be shown to follow from the Lagrangians $L_{\pm}$.
Indeed, the Lagrangian (\ref{ln3}) is already in the first order form
. Thus  we can read off the Hamiltonian directly
\begin{equation}
{\cal{H}_+} = \frac{\kappa}{2} \left(x_1^2 - x_2^2 \right)\label{hn1}
\end{equation}
with the symplectic algebra
\begin{equation}
\{x_i, x_j\} = -\frac{i}{g}\epsilon_{ij}\label{bn1}
\end {equation}
From (\ref{bn1}) we find that $igx_1$ is canonically conjugate to $x_2$.
Now by a canonical transformation to the set $(x,p_x)$ defined by
\begin{eqnarray}
x_1 = -\frac{i}{\sqrt{-\kappa}}p_x\hspace{.3cm};\hspace{.3cm}
x_2 = \frac{\sqrt{-\kappa}}{g}x\label{tn2}
\end{eqnarray}
the Hamiltonian (\ref{hn1}) becomes
\begin{equation}
{\cal{H}_+} = \left(\frac{p_x^2}{2} + \frac{\omega_+^2 x^2}{2} \right)\label{hn2}
\end{equation}
where we have used equations (23), (\ref{v}) and (\ref{freq}).
The above Hamiltonian coincides with $H_+$ of (\ref{dh}).
Similarly we can derive $H_-$ from $L_-$.
The correspondence between the Lagrangian and Hamiltonian formulations is
thus complete.

A question may arise regarding the interpretation of the complex Hamiltonians  $H_{\pm}$ found in the constituent pieces. The first point to note is that they satisfy
\begin{equation}
H_{\pm}^{\dagger} = H_{\mp}\label{hc}
\end{equation}
This hermitian conjugation proprty corresponds to the time reversal operation
that connects the doubled degrees of freedom of the closed theory.
Also, this property manifestly ensures the hermiticity of the complete 
Hamiltonian (\ref{hd}). 

Although $H_\pm$ are not hermitian, they are pseudo - hermitian \cite{AM},
\begin{equation}  
H_{\pm}^{\dagger} = \eta H_{\pm} \eta^{-1}\label{crucial}
\end{equation}
where $\eta$ is the PT operator. 
  Such Hamiltonians have occured in the study of 
PT-symmetric quantum mechanics \cite{B}, in minisuperspace 
quantum cosmology and other constructions\cite{AM}.  
To prove the condition (\ref{crucial}) note that
\begin{equation}
\eta x_i \eta^{-1} = g_{ij}x_j,\hspace{.3cm} \eta p_i \eta^{-1} =  -g_{ij}p_j
\end{equation}

The Hamiltonians $H_{\pm}$, given by (\ref{dh}), are of the form
\begin{equation}
H = \frac{p^2}{2} + \frac{\omega^2}{2}x^2\label{n1}
\end{equation}
where $x$ and $p$ are non - hermitian and $\omega$ is a complex number. Under
$\eta = PT$ the operators $x$ and $p$ transform as
\begin{equation}
\eta x \eta^{-1} = x^{\dagger}\hspace{.5cm}and \hspace{.5cm}\eta p \eta^{-1} =
-p^{\dagger}\label{n2}
\end{equation}
Now  define
\begin{equation}
a = \sqrt{\omega\over 2}\left( x + \frac{ip}{\omega}\right)\label{4.4}
\end{equation}
and
\begin{equation}
\tilde{a} = \eta^{-1} a^{\dagger}\eta = \sqrt{\frac{\omega}{2}}\left(x - \frac{ip}{\omega}\right)\label{4.5}
\end{equation}
Here $\tilde{a}$ is aptly called the pseudo - hermitian adjoint of $a$ with respect to
$\eta$. Writing     
\begin{equation}
N = \tilde{a}a\label{4.6}
\end {equation}
we get
\begin {equation}
H = \omega\left(N + \frac{1}{2}\right)\label{4.7}
\end{equation}
From the basic commutators between the canonical variables $x$ and $p$ 
 it is easy to derive that    
\begin{eqnarray}
\left[N,a\right] = -a\nonumber\\
\left[N,\tilde{a}\right] = \tilde{a}\label{4.11}
\end{eqnarray}
Also
\begin{equation}
\eta^{-1}N^{\dagger}\eta = N\label{4.12}
\end{equation}
Assume that we can construct a complete bidimensional eigenbasis 
$\{|\psi_n>,|\phi_n>\}$ diagonalising N
\begin{eqnarray}
N|\psi_n> & =& n|\psi_n>\nonumber\\
N^{\dagger}|\phi_n> & =& n^*|\phi_n>\nonumber\\
<\phi_n|\psi_m> & =& \delta_{nm}, \nonumber\\
\Sigma |\phi_n><\psi_n| & =& \Sigma |\psi_n><\phi_n| = 1\label{4.13}
\end{eqnarray}
Due to (\ref{4.7}) this is also the eigenbasis of the Hamiltonian. Now
using the commutation relations it can be shown that
\begin{equation}
N\left(a|\psi_n>\right) = (n - 1)a|\psi_n>\label{4.14}
\end{equation}
Hence we can write
\begin{equation}
a|\psi_n> = c|\psi_{n-1}>\label{4.15}
\end{equation}
where c is some c - number. Similarly
\begin{equation}
<\phi_n|\tilde{a} = d<\phi_{n-1}|\label{4.16}
\end{equation}
The pseudo - hermiticity of N can be exploited to relate $\eta|\phi_n>$ with $|\psi_n>$ because
\begin{equation}
N\eta|\phi_n> = n\eta|\phi_n>\label{e}
\end{equation}
Using the first equation of (\ref{4.13}) we find,
upto a phase, the following identification,
\begin{equation}
\eta|\phi_n> = |\psi_n>\label{4.17}
\end{equation}The correspondence (\ref{4.17}) enables us to reach a crucial result
\begin{equation}
<\phi_n|\tilde{a}|\psi_{n-1}> = <\phi_{n-1}|a|\psi_n>^*\label{4.18}
\end{equation}
which, along with (\ref{4.15}) and (\ref{4.16}) gives
\begin{equation}
d = c^*\label{4.19}
\end{equation}
The last result can be used to show that
\begin{equation}
n = |c|^2\label{4.20}
\end{equation}
We find that the eigenvalues of N are real and positive. 
We can also argue that it is integral otherwise repeated application of
$a$ would yield negative eigenvalue of N. There, thus, exists a state 
$|0>$ which is annihilated by $a$
\begin{equation}
a|0> = 0\label{4.21}
\end{equation}
Due to (\ref{4.7}) this state is the ground state of the Hamiltonian. From the ground state $|0>$ one can develop
all the higher energy states by repeated application of $\tilde{a}$.

  From the above solution of the eigenvalue problem of
 (\ref{n1}) we can build the physical
states of the composite system by forming direct product.
, Observe that due to (\ref{hc}) the eigenbasis
of $H_-$ will be $\{|\phi_n>,|\psi_n>\}$ if the eigenbasis of $H_+$ is
 $\{|\psi_n>,|\phi_n>\}$.

Any formulation of the d.h.o. is based on the direct \cite{Ba} or indirect
representation \cite{FT,C,BGPV}. The direct representation leads to lagrangians having an
explicit time dependence; hence these are not very popular. The indirect
representation avoids this problem by a doubling of the degrees of freedom.
It is called indirect because, taking the composite Lagrangian
 and varying one degree of freedom yields the equation of motion
 for the other degree
(see (\ref{n}) and its relevant equations of motion). The usual composite
Lagrangian, by construction, is two dimensional. It incorporates both
forward and backward time propagations. Individual one dimensional
Lagrangians displaying these properties were non - existant.

 The new point in our paper is that we have provided explicit one dimensional
lagrangians (equations (\ref{ln1}) and (\ref{ln2})) that characterise both
forward and backward time propagations. Although structurally these
Lagrangians look two dimensional, the symplectic structure effectively
reduces one dimension. Moreover we showed that a combination of these
Lagrangians led to (\ref{L2}). In this sense these lagrangians are more
fundamental. Also they cannot be obtained by taking the simple $m \to 0$
limit of (\ref{L2}). In the
 region of the parameter
space which corresponds to the damped oscillatory motion, the parameters
of the constituent Lagrangians were complex valued.
Also, these Lagrangians were complex conjugates of one another. Because
of this property,
 the resulting Hamiltonians
were complex valued, satisfying the requirements of pseudohermiticity. This
pseudohermiticity was exploited to diagonalize the individual Hamiltonians.
Based on this, an alternative quantization of the damped harmonic oscillator
was indicated.

{\bf Acknowledgement}

 One of the authors (P.M.) would like to thank Prof. 
S. Dattagupta, Director, S. N. Bose National Center for Basic Sciences,
for providing the necessary facilities to work as a 
Visiting Associate (U. G. C.). He also wishes to thank
the U. G. C. for the award of Visiting Associateship.

\end{document}